\documentclass[aps,prl,twocolumn,groupedaddress,showpacs]{revtex4}
\usepackage{graphicx}
\usepackage{dcolumn}
\usepackage{bm}
\begin{document}

\title{Relativistic-covariant Bohmian mechanics with proper foliation}

\author{Hrvoje Nikoli\'c}
\affiliation{Theoretical Physics Division, Rudjer Bo\v{s}kovi\'{c}
Institute,
P.O.B. 180, HR-10002 Zagreb, Croatia.}
\email{hrvoje@thphys.irb.hr}

\date{\today}

\begin{abstract}
In classical relativistic mechanics, a ``preferred'' proper direction in spacetime for each particle
is determined by the direction of its 4-momentum. Analogously, for each quantum particle
we find a local direction uniquely determined by the many-particle wave function,
which for each particle defines the proper foliation of spacetime.
This can be used to formulate a relativistic-covariant version 
of Bohmian mechanics, with equivariant probability density on proper hypersurfaces. 
\end{abstract}


\maketitle

Bohmian mechanics \cite{bohm} is a formulation of 
quantum mechanics in terms of 
deterministic particle trajectories,
with valuable interpretational \cite{book-bohm,book-hol,book-durr},
practical \cite{wyatt-prl,wyatt-book,apl-book}, 
and weakly measurable \cite{weak1,weak2,weak-science,weak4} aspects. 
A mayor remaining technical and conceptual challenge 
for Bohmian mechanics is to reconcile its explicit nonlocality with 
the theory of relativity.

A promissing approach 
to relativistic Bohmian mechanics is to formulate it in a manifestly
covariant form with the aid of an additional local unit 4-vector 
$N^{\mu}$ \cite{durr99,tumdis,nik-prob}.
This additional structure in the theory defines a 
``preferred" foliation of spacetime -- the foliation for which 
hypersurfaces are orthogonal to $N^{\mu}$. 
The problem with this approach is that the theory in its current form 
does not specify how to choose $N^{\mu}$. 

In the present paper we find the natural choice of $N^{\mu}$
for each particle uniquely determined by the many-particle wave function of the system.
(The possibility that $N^{\mu}$ could be determined by the state of the system was also
suggested in \cite{durr99,goldstein-zanghi}, but the specific proposals there were not 
fully satisfying \cite{goldstein}.)
In this way the ``additional" structure is not 
additional at all, but is already encoded in the wave function itself,
used also to calculate probability
densities and Bohmian particle velocities.

The basic physical idea is very simple. In empty spacetime with
Minkowski metric $g_{\mu\nu}$ (we use the signature $(+---)$
and the units $c=1$) there is no any preferred direction in spacetime.
However, this is no longer true when matter is present. In particular,
if there are $n$ classical particles at the spacetime positions
$x_a=\{ x_a^{\mu} \}$, $a=1,\ldots,n$, then the 
4-momentum $k_a^{\mu}$ of each particle defines a preferred direction
in spacetime at the position $x_a$. Each such direction defines a local
3-dimensional patch orthogonal to $k_a^{\mu}$,
corresponding to a local proper coordinate frame in which the particle is at rest.

To get a feeling how a quantum analogue of it may look like,
consider a many-time wave function of the form
\begin{equation}\label{pf1}
 \psi(x_1,\ldots,x_n)=\psi_1(x_1)\cdots \psi_n(x_n) ,
\end{equation}
where $\psi_a(x_a)=e^{-ik_{a\mu}x_a^{\mu}}$ is a momentum-eigenstate
plane wave. 
(The Einstein convention of summation over repeated indices refers
only to vector indices $\mu$, while the summation over the particle labels $a$
is to be performed only when the summation $\sum_a$ is indicated explicitly.)
The phase of the wave function is
$
S(x_1,\ldots,x_n)=\sum_a S_a(x_a) 
$
with $S_a(x_a) =-k_{a\mu}x_a^{\mu}$, so the vector
\begin{equation}\label{pf3}
 f_{a\mu}(x_a) \equiv -\partial_{a\mu}S(x_1,\ldots,x_n) = k_{a\mu} 
\end{equation}
defines the proper direction for the $a$'th particle {\em everywhere} in spacetime
(not merely ``at the position of the particle'', since the wave function
by itself does not determine a particle position).

In general, however, the wave function does not have a product form (\ref{pf1}),  
so the simple definition of the proper direction-vector in (\ref{pf3}) should be generalized
to a mathematically more sophisticated expression. In this paper we find 
such a more sophisticated expression generalizing (\ref{pf3}). After that we use it
to formulate Bohmian mechanics in a relativistic covariant form,
by generalizing the results of \cite{durr99,tumdis,nik-prob}
to include foliations which depend on the particle and are not spacelike everywhere.

Consider first particles without spin. The wave function $\psi(x_1,\ldots,x_n)$ satisfies $n$
Klein-Gordon equations
\begin{equation}\label{KG}
 (\partial^{\mu_a}\partial_{\mu_a}+m_a^2)\psi=0,
\end{equation}
one for each $a$, where $\partial_{\mu_a}\equiv \partial/\partial x^{\mu_a}_a$.
The crucial quantity calculated from $\psi$, from which everything else will be expressed, 
is the $n$-vector
\begin{equation}\label{curnold}
j_{\mu_1\ldots\mu_n}(x_1,\ldots,x_n)= \psi^*
\Gamma_{\mu_1} \cdots  \Gamma_{\mu_n}\psi ,
\end{equation}
where
$
 \Gamma_{\mu}\equiv \frac{i}{2} \stackrel{\leftrightarrow\;}{\partial_{\mu}}
$,
and $A\!\stackrel{\leftrightarrow\;}{\partial_{\mu}}\!B \equiv A (\partial_{\mu} B)-(\partial_{\mu} A) B$.  
Due to (\ref{KG}), the $n$-vector (\ref{curnold}) satisfies $n$ conservation equations
\begin{equation}\label{consn}
 \partial_{\mu_a}j^{\mu_1\ldots \mu_a \ldots \mu_n}=0 ,
\end{equation}
one for each $x_a$.

Now let $\Sigma_1,\dots,\Sigma_n$ be a collection of $n$ arbitrary hypersurfaces.
They do not need to be spacelike everywhere \cite{nikfol}, but we choose them
to be spacelike at infinity. The covariant measure of the 3-volume on $\Sigma_a$ is
\begin{equation}
dS^{\mu_a}=d^3x_a |g_a^{(3)}|^{1/2} n^{\mu_a} ,
\end{equation}
where $n^{\mu_a}(x_a)$ is the unit vector normal to $\Sigma_a$ and
$g_a^{(3)}(x_a)$ is the determinant of the induced metric on $\Sigma_a$.
For definiteness, $n^{\mu_a}$ is oriented such that it is future-oriented at infinity
where $\Sigma_a$ is spacelike. 
We take $\psi$ to be a superposition of positive-frequency solutions of (\ref{KG})
and normalize it such that the $n$-particle Klein-Gordon scalar product 
$(\psi,\psi)$ is equal to 1:
\begin{eqnarray}\label{KGn}
(\psi,\psi) & \equiv & 
\displaystyle\int_{\Sigma_1} dS^{\mu_1} \cdots \int_{\Sigma_n} dS^{\mu_n} \,
j_{\mu_1 \ldots\mu_n}
\nonumber \\
& = & 
\displaystyle\int_{\Sigma_1} d^3x_1 \cdots \int_{\Sigma_n} d^3x_n \,
\tilde{n}^{\mu_1} \cdots \tilde{n}^{\mu_n} j_{\mu_1 \ldots\mu_n}
\nonumber \\
& = & 1 ,
\end{eqnarray}
where
\begin{equation}\label{ntilde}
 \tilde{n}^{\mu_a}=|g_a^{(3)}|^{1/2} n^{\mu_a} ,
\end{equation}
and the tilde above $n^{\mu_a}$ denotes that $\tilde{n}^{\mu_a}$ transforms as a vector density.
The unit normal vector $n^{\mu_a}(x_a)$ is well-defined at points $x_a$
at which the hypersurface is spacelike or timelike. At points at which
it is null the quantities $n^{\mu_a}(x_a)$ and $|g_a^{(3)}(x_a)|^{1/2}$
are ill-defined, but their product (\ref{ntilde}) is well-defined everywhere \cite{nikfol}.
From (\ref{consn}) and the Gauss theorem one can see that
(\ref{KGn}) does not depend on the choice of hypersurfaces $\Sigma_1,\dots,\Sigma_n$.

The $n$-vector (\ref{curnold}) uniquely defines $n$ 1-particle currents
$j_{\mu_a}(x_a)$ by omitting the integration over $dS^{\mu_a}$ in  (\ref{KGn}).
For example, for $a=1$
\begin{equation}\label{eq6}
j_{\mu_1}(x_1)=
\int_{\Sigma_2} dS^{\mu_2} \cdots \int_{\Sigma_n} dS^{\mu_n} \,
j_{\mu_1 \mu_2\ldots\mu_n}(x_1,\ldots,x_n) ,
\end{equation}
and similarly for other $a$. Just like (\ref{KGn}), the current
(\ref{eq6}) also does not depend on the choice of hypersurfaces $\Sigma_1,\dots,\Sigma_n$.

In particular, for the product wave function as in (\ref{pf1}), one finds that
$j_{\mu_a}(x_a) \propto k_{\mu_a}$ (where $k_{\mu_a}\equiv k_{a\mu}$ 
and the constant of proportionality is irrelevant), so comparison 
with (\ref{pf3}) demonstrates that $j_{\mu_a}(x_a)$ could determine
the proper direction for the $a$'th particle. However, what we need is a vector
field $f_{\mu_a}(x_a)$ which defines a unique proper foliation of spacetime 
for the $a$'th particle, such that $f_{\mu_a}(x_a)$ is everywhere normal 
to the  proper-foliation hypersurfaces. One cannot simply take 
$f_{\mu_a}(x_a)$ to be equal to $j_{\mu_a}(x_a)$, because, in general,
for an arbitrary $j_{\mu}(x)$ there is no foliation with hypersurfaces
everywhere normal to $j_{\mu}(x)$. Instead, from a given $j_{\mu}(x)$
one needs to extract the appropriate $f_{\mu}(x)$ which does define
the foliation with hypersurfaces everywhere normal to $f_{\mu}(x)$.

The extraction of such $f_{\mu}(x)$ from a given $j_{\mu}(x)$
is a general mathematical problem. The solution, indicated also in \cite{durr99},
is as follows. A sufficient condition for $f_{\mu}(x)$ to define
a unique foliation is that it can be written as $f_{\mu}(x)=\partial_{\mu}\phi(x)$
for some function $\phi(x)$ \cite{poisson}.
Since $\partial_{\mu}\phi(x)$ is normal to the hypersurfaces, it follows that
\begin{equation}\label{phi}
 \phi(x) = \int dx^{\mu}\partial_{\mu}\phi(x) = \int dx^{\mu}f_{\mu}(x)  
\end{equation}
is constant on any hypersurface normal to $f_{\mu}(x)$.
The condition $f_{\mu}(x)=\partial_{\mu}\phi(x)$ implies
\begin{equation}\label{curle}
 \partial_{\nu}f_{\mu}(x)-\partial_{\mu}f_{\nu}(x)=0 .
\end{equation}
So, to extract the $f_{\mu}(x)$ satisfying (\ref{curle}) from given  $j_{\mu}(x)$,
we write $j_{\mu}(x)$ in terms of Fourier transforms
\begin{equation}\label{pf11}
 j_{\mu}(x)=\int \frac{d^4k}{(2\pi)^4} \, \hat{j_{\mu}}(k) e^{-ik\cdot x} ,
\end{equation}
\begin{equation}\label{pf12}
 \hat{j_{\mu}}(k)=\int d^4x' \, j_{\mu}(x') e^{ik\cdot x'} ,
\end{equation}
where $k\cdot x \equiv k_{\alpha} x^{\alpha}$.
Then $f_{\mu}(x)$ is determined by
\begin{equation}\label{pf13}
 f_{\mu}(x)=\int \frac{d^4k}{(2\pi)^4} \, \hat{f_{\mu}}(k) e^{-ik\cdot x} ,
\end{equation}
where
\begin{equation}\label{pf14}
 \hat{f_{\mu}}(k) = k_{\mu} \frac{\hat{j_{\alpha}}(k) k^{\alpha}} {k\cdot k} .
\end{equation}
Indeed, from (\ref{pf13}) with (\ref{pf14}) one easily finds that (\ref{pf13})
satisfies (\ref{curle}). 
In particular, if $j_{\mu}(x)=\partial_{\mu}\phi(x)$, one can check explicitly that the 
procedure (\ref{pf11})-(\ref{pf14}) gives $f_{\mu}(x)=j_{\mu}(x)$.
This shows that (\ref{pf13}) with (\ref{pf14}) extracts the foliation-defining part $f_{\mu}(x)$
of given $j_{\mu}(x)$.

Now we can turn back to physics. From $j_{\mu_a}(x_a)$ for each particle we extract 
$f_{\mu_a}(x_a)$ by the procedure above. This defines the proper foliation for each particle,
with the unit vector 
$N^{\mu_a}(x_a)= f^{\mu_a}/\sqrt{|f^{\nu_a}f_{\nu_a}|}$
normal to the hypersurfaces of proper foliation. Even though $N^{\mu_a}$ is ill-defined at points at which
$f^{\mu_a}$ is null, the direction of $N^{\mu_a}$ is well-defined, which for our purposes will turn out to be sufficient.
 
Note that, in general, 
$N^{\mu_a}(x)\equiv N_a^{\mu}(x)$ depends on $a$. 
If $\psi(x_1,\ldots,x_n)$ is symmetric or antisymmetric under the exchange 
of all $x_{a'}$, then $N_a^{\mu}(x)$ is the same for all $a$. 
But in general, the wave function may be neither symmetric nor antisymmetric.
In particular, when the masses $m_a$ in (\ref{KG}) depend on $a$, then
the particles are not identical, in which case there is no physical reason
to expect symmetry or antisymmetry of the wave function.

Now when we are equipped with unit $N^{\mu_a}(x_a)$ 
directed as $f^{\mu_a}(x_a)$ given by (\ref{pf13}),
we can formulate Bohmian mechanics in a relativistic-covariant form.
We introduce nonlocal vector fields $V_{\mu_a}(x_1,\ldots,x_n)$ by contracting
$j_{\mu_1\ldots\mu_n}$ with $(n-1)$ normals $N^{\mu_{a'}}$, 
$a'\neq a$.  For example, for $a=1$
\begin{eqnarray}\label{pfV}
V_{\mu_1}(x_1,\ldots,x_n) & = & j_{\mu_1\mu_2\ldots\mu_n}(x_1,\ldots,x_n)
\nonumber \\
& & N^{\mu_2}(x_2)\cdots N^{\mu_n}(x_n) ,
\end{eqnarray}
and similarly for other $a$. 
Even if the norm of $V_{\mu_a}$ is not well-defined at points at which 
$N^{\mu_{a'}}$ is null, the direction of $V_{\mu_a}$ is well-defined everywhere.
Therefore, it is consistent to postulate that the Bohmian particle trajectories are integral curves
of $V^{\mu_a}$. Such trajectories 
satisfy a covariant equivariance equation on proper hypersurfaces, which we now prove.

The proof rests on two crucial observations.
First, (\ref{consn}) implies $\partial_{\mu_a} V^{\mu_a} = 0$,
which we write in the form covariant under general coordinate transformation as
\begin{equation}\label{pf17}
\nabla_{\mu_a} V^{\mu_a} = 0,
\end{equation}
where $\nabla_{\mu_a}$ is the covariant derivative.
Second, (\ref{pfV}) implies that $N^{\mu_a}V_{\mu_a}$
does not depend on $a$, so that we have
\begin{equation}\label{pf18}
 N^{\mu_1}V_{\mu_1}=\cdots =N^{\mu_n}V_{\mu_n}= \rho ,
\end{equation}
where $\rho(x_1,\ldots,x_n)$ is defined as
\begin{equation}\label{pf19}
\rho =  j_{\mu_1\ldots\mu_n} N^{\mu_1}\cdots N^{\mu_n} .
\end{equation}

To prove the equivariance explicitly, we use the fact that
any vector $A^{\mu_a}$ can be decomposed as
\begin{equation}\label{decomp}
 A^{\mu_a}=A_{\parallel}^{\mu_a}+A_{\perp}^{\mu_a} ,
\end{equation}
where $A_{\parallel}^{\mu_a}$ is parallel with $N^{\mu_a}$, while 
$A_{\perp}^{\mu_a}$ is normal to $N^{\mu_a}$ (i.e., parallel 
with the proper hypersurface). 
More explicitly, 
\begin{equation}\label{paral}
 A_{\parallel}^{\mu_a}=N^{\mu_a}
\frac{N^{\alpha_a}A_{\alpha_a}}{N^{\nu_a} N_{\nu_a}}
=u^{\mu_a} N^{\alpha_a}A_{\alpha_a} ,
\end{equation}
where
$
 u^{\mu_a}=N^{\mu_a}/{N^{\nu_a} N_{\nu_a}} 
$.
Eqs.~(\ref{paral}) and (\ref{decomp}) give $A_{\parallel}^{\mu_a} B_{\perp\mu_a}=0$
for any two vectors $A^{\mu_a}$ and $B^{\mu_a}$. Therefore 
(\ref{pf17}) can be decomposed as
\begin{equation}\label{pf17decomp}
\nabla_{\parallel\mu_a} V_{\parallel}^{\mu_a} + \nabla_{\perp\mu_a} V_{\perp}^{\mu_a}= 0.
\end{equation}
Using (\ref{paral}), the first term in (\ref{pf17decomp}) can be written as
$
\nabla_{\parallel\mu_a} V_{\parallel}^{\mu_a} = u_{\mu_a}N^{\alpha_a}\nabla_{\alpha_a}
(u^{\mu_a}N^{\beta_a}V_ {\beta_a}) 
$.
Here $N^{\beta_a}V_ {\beta_a}=\rho$ due to (\ref{pf18}), so
$
\nabla_{\parallel\mu_a} V_{\parallel}^{\mu_a} = u_{\mu_a}u^{\mu_a}
N^{\alpha_a}\nabla_{\alpha_a} \rho + \rho \delta_a,
$
where $\delta_a=N^{\alpha_a}u_{\mu_a}\nabla_{\alpha_a}u^{\mu_a}
=\frac{1}{2}N^{\alpha_a}\nabla_{\alpha_a}(u^{\mu_a}u_{\mu_a})$ 
is proportional to a Dirac $\delta$-function vanishing everywhere except at points at
which the unit norm $u^{\mu_a}u_{\mu_a}$ changes sign. Such a singular term
$\rho \delta_a$ appears also in the second term of (\ref{pf17decomp}) with the opposite sign,
so the singular terms cancel up in (\ref{pf17decomp}). Thus it is consistent to redefine both terms in
(\ref{pf17decomp}) so that the singular term is subtracted from each of them. As a result,
with such a redefinition we have 
\begin{equation}\label{nablaV}
 \nabla_{\parallel\mu_a} V_{\parallel}^{\mu_a} = 
u_{\mu_a}u^{\mu_a} N^{\alpha_a}\nabla_{\alpha_a} \rho .
\end{equation}
Next we parameterize the integral curves of $V^{\mu_a}$ as $X^{\mu_a}(s)$
with a scalar parameter $s$ increasing along the curves, so that 
\begin{equation}\label{bohmtraj}
 \frac{dX^{\mu_a}(s)}{ds}=v^{\mu_a}(X_1(s),\ldots,X_n(s))  
\end{equation}
where
\begin{equation}\label{v}
 v^{\mu_a} \equiv \frac{V^{\mu_a}}{|\rho|}  \; \Rightarrow \;
N^{\mu_a}v_{\mu_a}={\rm sign}\, \rho ,
\end{equation}
the last equality is a consequence of (\ref{pf18}), 
and ${\rm sign}\, \rho =\rho/|\rho|$.
In local coordinates $x^{\mu_a}=(x^{0_a},{\bf x}_a)$ in which
$N^{\mu_a}=(1,0,0,0)$, one can introduce the quantity 
$\rho({\bf x}_1,\ldots,{\bf x}_n,s) \equiv \rho(X^{0_1}(s),{\bf x}_1,\ldots, X^{0_n}(s),{\bf x}_n)$,
implying
\begin{equation}\label{rhos}
 \frac{\partial\rho}{\partial s}=\sum_{a=1}^{n} \frac{dX^{0_a}}{ds}
\frac{\partial\rho}{\partial x^{0_a}} 
=\sum_{a=1}^{n} v^{0_a}\partial_{0_a} \rho .
\end{equation}
The covariant version of (\ref{rhos}), valid everywhere for any $N^{\mu_a}(x_a)$, is
\begin{eqnarray}\label{rhoscov}
 \frac{\partial\rho}{\partial s} & = &
\sum_{a=1}^{n} v_{\parallel}^{\mu_a}\nabla_{\parallel\mu_a} \rho
=\sum_{a=1}^{n} u^{\mu_a} N^{\alpha_a} v_{\alpha_a} u_{\mu_a} N^{\beta_a}\nabla_ {\beta_a} \rho
\nonumber \\
& = & {\rm sign}\, \rho \sum_{a=1}^{n} u^{\mu_a} u_{\mu_a} N^{\beta_a}\nabla_ {\beta_a} \rho
\nonumber \\
& = & 
{\rm sign}\, \rho \sum_{a=1}^{n} \nabla_{\parallel\mu_a} V_{\parallel}^{\mu_a} ,
\end{eqnarray}
where (\ref{v}) and (\ref{nablaV}) were used in the second and third line, respectively.
Therefore, by summing (\ref{pf17decomp}) over $a$ and using (\ref{rhoscov}) and (\ref{v}),
we finally get 
\begin{equation}\label{equivar}
 \frac{\partial |\rho|}{\partial s}+ \sum_{a=1}^{n}\nabla_{\perp\mu_a} (|\rho| v_{\perp}^{\mu_a})= 0 .
\end{equation}
This can be recognized as the covariant equivariance equation for the
probability ``density'' $|\rho|$. More precisely, the probability density on proper hypersurfaces
transforming as a scalar density is
\begin{equation}\label{probdens}
\tilde{p}(x_1,\ldots,x_n) = |\tilde{\rho}(x_1,\ldots,x_n)| ,
\end{equation}
where
$
\tilde{\rho} =  j_{\mu_1\ldots\mu_n} \tilde{N}^{\mu_1}\cdots \tilde{N}^{\mu_n} 
$
is well-defined even at points at which a proper hypersurface is null
(see Eq.~(\ref{ntilde}) and the discussion of it).  

The parameter $s$ can be used to parameterize the proper
hypersurfaces as $\Sigma_a(s)$. Namely, each proper hypersurface $\Sigma_a$
is defined by a value $\phi_a$ constant on the hypersurface, 
where $\phi_a$ is a function of $s$ determined by (\ref{phi}) and 
(\ref{bohmtraj}). Explicitly, this function is 
\begin{equation}
 \phi_a(s)=\int_{0}^{s} ds \, \frac{dX^{\mu_a}}{ds} \, f_{\mu_a}
= \int_{0}^{s} ds \, v^{\mu_a} f_{\mu_a} ,
\end{equation}
where the integrals are evaluated along the trajectories (\ref{bohmtraj}).
Hence, if a statistical ensemble of particles with velocities (\ref{bohmtraj}) has the probability
distribution (\ref{probdens}) at some initial collection of proper hypersurfaces
$\Sigma_1(s=0),\ldots, \Sigma_n(s=0)$, then (\ref{equivar}) implies that the ensemble
has the distribution (\ref{probdens}) at $\Sigma_1(s),\ldots, \Sigma_n(s)$
for any $s$,
which finishes the proof of equivariance.

Concerning the probability density (\ref{probdens}), one additional comment is in order.
In general, $\tilde{N}^{\mu_1} \cdots \tilde{N}^{\mu_n} j_{\mu_1\ldots\mu_n}$
may be negative at some parts of proper hypersurfaces. Thus, the comparison with
(\ref{KGn}) implies 
\begin{equation}\label{>1}
 \displaystyle\int_{\Sigma_1} d^3x_1 \cdots \int_{\Sigma_n} d^3x_n \, \tilde{p} \geq 1.
\end{equation}
The case $>1$ has a simple physical origin \cite{tumdis,nik-prob}.
This happens when the congruence of all particle trajectories satisfying (\ref{bohmtraj})
is such that some trajectories cross some proper hypersurface $\Sigma_a$ more than ones. 
If one takes truncated hypersurfaces $\Sigma'_a \subset \Sigma_a$ such that each
trajectory $X_a(s)$ crosses $\Sigma'_a$ ones and only ones, then the integral 
(\ref{>1}) (with the integration-region replacements $\Sigma_a \rightarrow \Sigma'_a$) 
is strictly equal to 1 \cite{nik-prob}.

Finally, let us generalize all this to the case of particles with spin. The only non-trivial issue
is to find a generalization of (\ref{curnold}), because once $j_{\mu_1\ldots\mu_n}$
with property (\ref{consn}) is known, the rest of the procedure is the same as 
for spinless particles above. The wave function $\psi_{l_1\ldots l_n}(x_1,\ldots, x_n)$
of $n$ particles with spin carries $n$ discrete spin indices $l_1,\ldots, l_n$.
Each component $\psi_{l_1\ldots l_n}(x_1,\ldots, x_n)$ with fixed values
of $l_1,\ldots, l_n$ satisfies the Klein-Gordon equations (\ref{KG}).
(For spin-$\frac{1}{2}$ and spin-$1$ Klein-Gordon equations
see, e.g., \cite{nikapl}.) 
Thus, the obvious generalization of (\ref{curnold}) satisfying (\ref{consn}) is
\begin{equation}\label{curnoldn}
j_{\mu_1\ldots\mu_n}= \psi^{\dagger}
\Gamma_{\mu_1} \cdots  \Gamma_{\mu_n}\psi ,
\end{equation}
where
$
 \psi^{\dagger} A \psi \equiv \sum_{l_1,\ldots, l_n} \psi^* _{l_1\ldots l_n} A \,
\psi _{l_1\ldots l_n}
$
for any object $A$ not carrying spin indices $l_1,\ldots, l_n$.

However, the case of spin-$\frac{1}{2}$ requires a more careful discussion.
This case has been studied in more detail in \cite{durr99,tumdis}, where instead of
(\ref{curnoldn}) a different choice has been proposed
\begin{equation}\label{curnoldn-dir}
j'_{\mu_1\ldots\mu_n}= \bar{\psi}
\gamma_{\mu_1} \cdots  \gamma_{\mu_n}\psi .
\end{equation}
Here $\gamma_{\mu_1} \cdots  \gamma_{\mu_n}$ is the direct product of $n$ Dirac
matrices and $\bar{\psi}=\psi^{\dagger}\gamma^{0_1}\cdots \gamma^{0_n}$. 
It has the advantage that $j'_{0_1\ldots 0_n}$ is positive definite
and (\ref{eq6}) is timelike everywhere, so (\ref{probdens}) can be taken without the absolute value.
The problem with (\ref{curnoldn-dir}) is that it cannot
be generalized to spin-$0$ and  spin-$1$, while (\ref{curnoldn}) works for any spin.

When (\ref{curnoldn}) is applied to spin-$\frac{1}{2}$, an additional clarification
is needed concerning the transformation properties of (\ref{curnoldn}).
From known transformation properties of spinors under Lorentz transformations
\cite{bd1}, one might naively conclude that (\ref{curnoldn}) does not transform as an
$n$-vector. However, this is not really true \cite{nikQFT,nikapl}. 
The standard spinor-transformation properties \cite{bd1} cannot be generalized to curved
spacetime, so for general purposes it is more convenient to redefine the transformation properties
of spinors and Dirac matrices such that $\psi$ transforms as a scalar and $\gamma_{\mu}$ 
as a vector under coordinate transformations \cite{weinberg,birdel}. Such a redefinition
of transformations does not alter the $n$-vector transformation properties of 
(\ref{curnoldn-dir}), but implies that (\ref{curnoldn}) also transforms as an $n$-vector.

To conclude, in this paper we have shown that Bohmian mechanics can be formulated
in a relativistic-covariant form. The central quantity calculated from the wave function 
is the conserved $n$-vector $j_{\mu_1\ldots\mu_n}$ from which one calculates
$j_{\mu_a}(x_a)$ given by (\ref{eq6}), the foliation-defining part $f_{\mu_a}(x_a)$
of which is given by (\ref{pf13})-(\ref{pf14}). This determines the proper foliation for each particle,
which, in turn, can be used to formulate Bohmian mechanics in a unique 
relativistic-covariant form by generalizing the methods developed earlier in \cite{durr99,tumdis,nik-prob}.


\end{document}